\documentclass[pra, 12 pt]{revtex4}
\usepackage[english]{babel}
\usepackage{amsmath}
\usepackage{epsfig}

\begin{document}

\title{AHARONOV--BOHM EFFECT, ELECTRODYNAMICS POSTULATES, AND LORENTZ CONDITION}

\author{V.B. Bobrov $^{\ast,\dagger}$, S.A. Trigger $^{\ast,\star}$, G.J.F.\; van Heijst $^{\star}$, P.P.J.M. Schram $^{\star}$}

\maketitle

\textbf{The problem of the relation between the Ahronov-Bohm effect and traditional postulates of electrodynamics, which claim that only electric and magnetic fields are observable, is resolved by denial of the statement about validity of the Maxwell equations for microscopic fields. We proceed from the idea that the Maxwell equations, as the generalization of experimental data, are valid only for averaged values. We show that microscopic electrodynamics should be based on postulation of the d'Alembert equations for four-vector of the electromagnetic field potential. The Lorentz condition is valid only for the averages and provides the implementation of the Maxwell equations for averages. This concept eliminates the problem of electromagnetic field quantization and provides the correctness of all known results of quantum electrodynamics. Therefore, the "virtuality" of the longitudinal and scalar photons has a formal mathematical character, conditioned by the Maxwell equations for averaged fields. The longitudinal and scalar photons provide not only the Coulomb interaction of charged particles, but also allow the electrical Aharonov-Bohm effect.}\\

Although the paper by Aharonov and Bohm [1] was published more than 50 years ago, various aspects of the Aharonov-Bohm  (AB)  effect still receive great attention up to present [2--10]. Most studies, including experimental researches, are devoted to the magnetic AB effect (see, e.g., [11]). \\
---------------------------------------------

$^{\ast}$ Joint\, Institute\, for\, High\, Temperatures, Russian\, Academy\,
of\, Sciences, Izhorskaya str. 13 build. 2, 125412 Moscow, Russia, $^\dagger$ National Research University "MPEI"\,,
Krasnokazarmennaya str. 14, 111250 Moscow,  Russia, $^{\star}$ Eindhoven  University of Technology, P.O. Box 513, MB 5600
Eindhoven, The Netherlands.\\

However, in recent years the interest in the electric AB effect has significantly increased (see [7, 12, 13] and references therein), which is due to different interpretations of the experiments [14,15].

Furthermore, taking into account the known analogy between electrostatic and gravitational interactions, the problem of the possible experimental validation of the AB effect in the gravitational field is posed in [16].

The problem of the electric AB effect concerns the basic concepts of the electromagnetic field gauge theory [17]. Since the magnetic field is completely defined by the transverse part of the vector potential, the magnetic AB effect is in fact independent of the gauge invariance of the electrodynamic equations [18]. At the same time, the electric AB effect feasibility creates an apparent contradiction with electrodynamics postulates.

One of the main arguments against the physical reality of the electromagnetic potential is the indeterminacy of its gauge choice. The physical reality of some variable means that its average value (at least for some ranges of the parameters and arguments on which this variable depends) can be single-valued by recreation out of some experiment. In most cases it is necessary to use a theory (always approximate) for the transition (or conversion) from a directly measured variable to the sought physically real variable. From a purely theoretical point of view we can consider some variable as physically real if it possesses a single-valued operator representation and carries as physical valued information. It is necessary to stress that there is a well-marked difference between the microscopical variables and their averaged values which only can be found form the experiments, if these variables are physically real. For shortness, the average values of the physically real variables  are referred to as observables.

As is known, the gauge (gradient) invariance is connected with the property of the Maxwell tensor for the electromagnetic field to remain unchanged when adding a four-dimensional gradient $\partial \chi/\partial x^\mu$ of an arbitrary scalar function $\chi$ of space--time coordinates is added to the four-vector of the electromagnetic potential. The requirement of relativistic invariance slightly narrows the arbitrariness in the choice of this function (the so-called Lorentz gauge). However, the use of the Lorentz gauge creates significant problems in electromagnetic field quantization, since the electromagnetic potential components become dependent on each other (see, e.g., [19]). But if only the electromagnetic field (i.e., electric and magnetic field strengths) is accepted as a physical reality, then the requirement for relativistic invariance of the potential seems to become excessive. Therefore, the Coulomb and other obviously non-covariant gauges are used in many studies.

Let us pose the following question; what is the basis of the statement about the physical reality of only specifically electromagnetic field strengths? It is easy to verify that the basis of this statement is the "tacit" postulate about the validity of Maxwell's equations for microscopic fields (see, e.g., [20]). However, the Maxwell equations, which are generalizations of available experimental data on the macroscopic behavior of systems of charged particles and correspond to experimentally confirmed concepts of the finite propagation velocity of the interaction between them, are valid only for averaged (or observed) values of dynamic variables (operators in quantum electrodynamics) entering these equations.

By virtue of the linearity of the Maxwell equations, the "tacit" assumption of the validity of these equations for dynamic variables naturally leads to their validity for average values as well. Strictly speaking, however, the converse is not necessarily valid. The demand of the fulfilment of this assumption follows neither from experimental data nor from the logical construction of the theory itself.

This means that we can require nothing more than that the equations of motion for microscopic fields would satisfy the Maxwell equations after the averaging procedure. Indeed, in attempting to construct microscopic electrodynamics, we cannot be restricted to the introduction of only electromagnetic field strengths, but we are "forced" to take into consideration the field potential to describe the field interaction with charged particles [20].
As is known, one of the key features of Maxwell equations is the absence of solutions to these equations, corresponding to longitudinal waves in vacuum, i.e., the absence of waves whose electric field vector is collinear to the wave vector. Nevertheless, longitudinal and scalar photons are used in quantum electrodynamics to describe the Coulomb interaction of charged particles
(see, e.g., [21]); these photons are considered as "nonphysical" (virtual), since, otherwise, the microscopic Maxwell equations will be invalid. However, as follows from the above consideration, there is no need for the requirement of the validity of the microscopic Maxwell equations.

The following question then arises; what can be proposed instead of microscopic the Maxwell equations? Quantum electrodynamics gives a factual answer to this question. Within the concept of the free electromagnetic field, it is quite natural to postulate that the microscopic four-vector potential of the electromagnetic field satisfies the d'Alembert equation [21]. In order that this equation would correspond to macroscopic Maxwell equations, it is necessary to require the validity of the Lorentz gauge condition, but only for average (observed) values of the four-potential of the electromagnetic field. However, this condition is completely consistent with the modern results of quantum electrodynamics from the viewpoint of the above mentioned difficulties in electromagnetic field quantization [19, 21]. In this case, the Lorentz condition for averages only means the choice of feasible quantum states, providing the independence of longitudinal and scalar photons in their microscopic description.

As a result, the fundamental difference in our interpretation of quantum electrodynamics is reduced to the statement that longitudinal and scalar photons are quite physical, rather than virtual. Their "nonphysicality" is reduced to nothing more than that they "are absent" in the macroscopic Maxwell equations for the free electromagnetic field. However, the Maxwell equations themselves are not unique from the viewpoint of the description of electromagnetic field manifestations.
In this sense, the analogy between the Klein--Gordon and Dirac equations is relevant: the former is used to determine Dirac matrices; however, the Dirac equation does not follow from the Klein--Gordon equation (see, e.g., [21]).

We also note that, instead of the d'Alembert equation for the four-vector potential of the electromagnetic field, we can postulate the corresponding Lagrangian for the system of charged particles and electromagnetic field, which was proposed by Fock and Podolsky as early as in 1932 [22]. Certainly, in this case, it should be kept in mind that it is not a single Lagrangian providing the validity of d'Alembert equation; however, the Fock--Podolsky Lagrangian corresponds to the "required" quantization procedure of the electromagnetic field and its interaction with charged particles, which themselves are described by the Dirac equation. Furthermore, the Lagrangian of the free electromagnetic field is no longer completely defined by the Maxwell tensor.

Then, how can we explain the fact that many results of applying gauge quantum electrodynamics lead to equivalent results using various gauges, first of all, the Lorentzian and Coulomb ones (see, e.g., [23])? This is due to the so-called "heuristic" quantization proposed by Feynman [24] (see [25] for more details). Calculating the radiative corrections to scattering, Feynman has paid attention to the fact that scattering amplitudes of elementary particles are independent of the reference frame and gauge choice. The independence of the reference frame came to be simply called the relativistic invariance, and the gauge choice became a formal procedure of choosing gauge-invariant field variables. However, the range of application of heuristic quantization is restricted only the problem of elementary particle scattering, where this quantization has arisen. This does exceed the problems of quantum electrodynamics (see, e.g., [26, 27]).

The essence of the concept formulated above leads to the following relations.
We proceed from the concept that the microscopic values of the vector potential ${\bf A}$ and scalar potential $\varphi$, forming the four-vector $A^\mu\equiv(\varphi, {\bf A})$, satisfy the d'Alembert equations
\begin{equation}
\frac{1}{c^2} \frac{\partial^2 {\bf A}}{\partial t^2}- \triangle {\bf A}  = \frac{4\pi {\bf j} }{c}\;, \qquad
\frac{1}{c^2} \frac{\partial^2 {\varphi}}{\partial t^2}- \triangle \varphi  = 4\pi\rho, \label{1a}
\end{equation}
or in covariant form
\begin{equation}
\textsc{D} A^\mu=4\pi j^\mu. \label{2a}
\end{equation}
where $j^\mu\equiv (\rho, {\bf j}/c)$ is the four-vector of the charge current and  $\textsc{D}$ is the d'Alembert operator
\begin{equation}
\textsc{D}\equiv\frac{1}{c^2} \frac{\partial^2}{\partial t^2}- \triangle\equiv\partial_\mu \partial^\mu, \qquad \partial_\mu=\left(\frac{\partial}{\partial c t}\,, \nabla\right),\;\partial^\mu=\left(\frac{\partial}{\partial c t}\,, -\nabla\right). \nonumber\\
\end{equation}
Due to linearity of the d'Alembert equations (1), (2) they are also valid for the average values $\langle A^\mu \rangle$\ and $\langle j^\mu \rangle$.
In the framework of quantum electrodynamics [21] the quantized free electromagnetic field should satisfy to the homogeneous d'Alembert equations (1),(2) (in absence of charges)
\begin{equation}
\textsc{D} A^\nu=0. \label{3a}
\end{equation}
However, for realization of the non-contradicted quantization of the free electromagnetic field it is necessary to fulfill the Lorentz condition for the averages  $\langle A^\mu\rangle$ (see in detail [7, 19, 21])
\begin{equation}
\partial_\mu \langle A^\mu\rangle=0. \label{4a}
\end{equation}
The essential point is the requirement that the Lorentz condition (4) should be satisfied not for the microscopic values of the four-vector $A^\mu$, as it is traditionally accepted in quantum electrodynamics in the so-called Lorentz calibration [20], but only for the average values $\langle A^\mu\rangle$.

This implies that the components of the microscopic four-vector $A^\mu$ can be considered independently in the quantization procedure of the free electromagnetic field. The Lorentz condition (4) leads only to limitations of the vectors of state of the electromagnetic field under consideration.
Since the condition (4) is evidently independent of whether charges are present or not, it can be postulated also in the presence of charged particles. It is then easy to see, taking the d'Alembert equations (1)-(2) for averages and using equation (4), the Maxwell equations for the average fields follow straightforwardly.

Taking into account that the average values of the electric and magnetic fields $\langle{\bf E}({\bf r},t)\rangle$ and $\langle{\bf H}({\bf r},t)\rangle$ are determined by
\begin{equation}
\langle{\bf E}\rangle=-\frac{1}{c} \frac{\partial \langle{\bf A}\rangle}{\partial t}-grad \langle\varphi\rangle,\; \; \langle{\bf H}\rangle=curl \langle{\bf A}\rangle,\label{5a}
\end{equation}
the Maxwell equations for the averaged $\langle A^\mu \rangle$ take the form
\begin{equation}
\textsc{D} \langle A^\mu\rangle -\partial^\mu \left(\partial_\alpha\langle A^\alpha \rangle\right)=4\pi \langle j^\mu \rangle. \label{6a}
\end{equation}
Here, as usual, the repeating indices mean summation. Using condition (4) we arrive at the d'Alembert equations for the average values $\langle A^\mu\rangle$
\begin{equation}
\textsc{D} \langle A^\mu\rangle=4\pi \langle j^\mu \rangle. \label{7a}
\end{equation}
These equations correspond to the averaged initial equations (1), (2) for the microscopical values of the potentials.

As a result, we come to the conclusion that if the d'Alembert equation for microscopical values of the four-vector potential of the electromagnetic field is accepted as the theoretical basis, and the Lorentz condition is valid only for its average (observed) value, we obtain all known results of quantum electrodynamics, but also confirm the possibility not only of the magnetic, but also the electric Aharonov--Bohm effect. In this case, Maxwell equations are valid only for average values of the electric and magnetic field.

In contrast to the Maxwell equations for average electric and magnetic strengths, the equations for the potentials permit solutions in the form of longitudinal waves (longitudinal and scalar photons), as well as for transversal ones. This result is crucial for quantum electrodynamics.

\section*{Acknowledgment}
This study was supported by Netherlands Organization for Scientific Research (NWO) and by joint grant of Russian Foundation for Basic Research and Ukraine National Academy of Science, projects No. 12-08-00600-a and No. 12-02-90433-Ukr-a. S.A. Trigger is thankful to NWO for support of his research by the individual grant in the years 2012-2013.
The authors are thankful to A.I. Ershkovich, A.M. Ignatov, A.A. Roukhadze and I.M. Sokolov for the fruitful discussions.\\

\section*{Authors contribution}
The theoretical concepts were developed and analyzed by V.B. Bobrov, S.A. Trigger, G.J.F. van Heijst, P.P.J.M. Schram. The authors contributed to discussion of the results and preparation of the manuscript.

\end{document}